\documentclass[a4paper,11pt]{article}
\usepackage{pos}
\usepackage{bm}

\title{Femtoscopy for exotic hadrons and nuclei}

\author*[a,b]{Tetsuo Hyodo}

\affiliation[a]{Department of Physics, Tokyo Metropolitan University, Hachioji 192-0397, Japan}

\affiliation[b]{RIKEN Interdisciplinary Theoretical and Mathematical Science Program (iTHEMS), Wako 351-0198, Japan}

\emailAdd{hyodo@tmu.ac.jp}

\abstract{In high energy collision experiments with multiple hadron productions, the momentum distribution of the measured hadron pair shows a correlation due to the final state hadron interactions and the quantum statistics. In the past, this femtoscopy technique has been developed to extract the information of the emission source from the momentum correlation functions. Recently, correlation function measurement is utilized also as a new method to determine the hadron interactions. In fact, the ALICE collaboration at LHC measures the correlation functions with various hadron pairs, for which the standard scattering experiment is difficult, providing remarkable progress in the study of the hadron interactions. In this contribution, we introduce the theoretical method to calculate the correlation functions, and present recent results on the study of the antikaon-nucleon interactions and hyperon-nucleus interactions.}

\FullConference{The XVIth Quark Confinement and the Hadron Spectrum Conference (QCHSC24)\\
 19-24 August, 2024\\
 Cairns Convention Centre, Cairns, Queensland, Australia\\}

\begin{document}
\maketitle

\section{Introduction}

The study of hadron-hadron interactions is an important subject for understanding the properties of hadrons. Initiated by Yukawa's meson theory for nuclear forces~\cite{Yukawa:1935xg}, this field has developed through extensive theoretical and experimental efforts. In recent years, advancements in accelerator experiments have led to the observation of numerous exotic hadrons~\cite{Hosaka:2016pey,ExHIC:2017smd,Guo:2017jvc,Brambilla:2019esw,Hyodo:2020czb,ParticleDataGroup:2024cfk}, prompting further investigation into their internal structure and production mechanisms. In particular, hadron-hadron interactions serve as the fundamental building blocks for studying hadronic molecules, which are loosely bound systems of hadrons, similar to conventional atomic nuclei~\cite{Guo:2017jvc}.

Traditional approaches to study hadron-hadron interactions rely on scattering experiments, which have been successful in establishing the nuclear force~\cite{Epelbaum:2008ga,Machleidt:2011zz}. However, due to the need for a stable target and a well-controlled beam, hadron systems suitable for scattering experiments are mostly limited to $YN$, $\pi N$, $KN$, and $\bar{K}N$. In fact, for interactions involving heavy hadrons containing $c$ and $b$ quarks, which are relevant to the study of exotic hadrons, scattering experiments are nearly impossible. Even for channels where scattering experiments are feasible, low-energy data are often not very precise due to the instability of the beam particles.

Recently, the femtoscopy technique has provided a novel method to probe hadronic interactions~\cite{ExHIC:2017smd,Fabbietti:2020bfg}. In contrast to scattering experiments, femtoscopy utilizes correlation functions extracted from high-energy collisions, allowing for high-precision studies of hadron-hadron interactions. In particular, experimental data are already available for multi-strangeness channels such as the $\Lambda\Lambda$~\cite{STAR:2014dcy,ALICE:2018ysd,ALICE:2019eol}, $N\Xi$~\cite{ALICE:2019hdt,ALICE:2020mfd}, and $N\Omega$~\cite{STAR:2018uho,ALICE:2020mfd} correlations. Moreover, correlation functions have also been measured in the charm sector, including the $D^{-}p$ pairs~\cite{ALICE:2022enj}, as well as the $D\pi$ and $DK$ pairs~\cite{ALICE:2024bhk}. These channels are almost impossible to access in traditional scattering experiments and provide important information on hadron interactions in conjunction with theoretical investigations~\cite{Morita:2014kza,Morita:2019rph,Kamiya:2021hdb}. Notably, correlation function measurements demonstrate excellent precision in low-energy correlations compared to traditional scattering experiments.

In the following, we first introduce the basic idea of the femtoscopy technique for investigating hadron-hadron interactions. Next, we present the results of the $K^{-}p$ correlation functions, which provide insights into the antikaon-nucleon interactions and the $\Lambda(1405)$ resonance. This study has been conducted as a joint effort between theoretical calculations and experimental analysis~\cite{ALICE:2019gcn,Kamiya:2021hdb,ALICE:2021szj,ALICE:2022yyh}. Finally, we show recent theoretical predictions for the correlation functions of hyperons and alpha particles~\cite{Jinno:2024tjh}.

\section{Theoretical concept}

We consider high-energy collision experiments, where a large number of hadrons are produced in a single collision. In particular, we focus on the collisions where the number of emitted particles is sufficiently large, allowing for a statistical treatment of particle production. In this case, the momentum correlation function for the particle 1 with momentum $\bm{p}_{1}$ and the particle 2 with momentum $\bm{p}_{2}$ can be defined as
\begin{align}
    C(\bm{q}) = \frac{N_{12}(\bm{p}_{1},\bm{p}_{2})}{N_{1}(\bm{p}_{1})N_{2}(\bm{p}_{2})}
    \label{eq:Cq}
\end{align}
where, $\bm{q}$ represents the relative momentum of the hadron pair, which is given in the center-of-mass frame of the hadron pair (pair rest frame) as $\bm{q} = \bm{p}_1 - \bm{p}_2$. The numerator of Eq.~\eqref{eq:Cq} corresponds to the yield of particle pairs measured simultaneously, denoted as $N_{12}$, while the denominator is the product of the individual yields of the same hadron species observed independently, $N_1 \cdot N_2$. Both quantities are normalized such that their momentum integrals sum to unity. If there were no final-state interactions between the hadrons, the correlation function for pairs of distinguishable hadrons would be $C(\bm{q}) = 1$ for all $\bm{q}$. For identical hadrons, only correlations arising from quantum statistics would be present. Therefore, by precisely measuring deviations from this reference correlation, the effects of hadron-hadron interactions can be extracted.

Theoretical calculation for the correlation function is usually performed by the Koonin-Pratt formula~\cite{Koonin:1977fh,Pratt:1986cc}. In fact, under several assumptions, the definition~\eqref{eq:Cq} can be rewritten as~\cite{Lisa:2005dd,ExHIC:2017smd,Murase:2024ssm} 
\begin{align}
   C(\bm{q}) = \int d^3r\ S(\bm{r})\ | \Psi^{(-)}_{\bm{q}}(\bm{r}) |^2,
   \label{eq:KPformula}
\end{align}
where $S(\bm{r})$ is the source function, representing the shape and size of the emission source of the hadron pair and $\Psi^{(-)}_{\bm{q}}(\bm{r})$ is the scattering wave function of the hadron pair in the relative coordinate $\bm{r}$ with an eigenmomentum $\bm{q}$, incorporating information on the hadron-hadron interaction. In conventional scattering theory, the $S$-matrix is obtained from the coefficient of the outgoing wave by normalizing the incident wave to unity. In contrast, for correlation function calculations, the wave function $\Psi^{(-)}$ is normalized such that the outgoing wave corresponding to the final-state hadron pair has a coefficient of unity.  

Equation~\eqref{eq:KPformula} implies two possible applications of the measurement of $C(\bm{q})$. If the two-body interaction is well studied and a reliable wave function $\Psi^{(-)}_{\bm{q}}(\bm{r})$ can be computed, measuring $C(\bm{q})$ enables the extraction of information about the hadron emission source $S(\bm{r})$~\cite{Goldhaber:1960sf,Lisa:2005dd}. This is the conventional usage of femtoscopy in the sense of performing microscopy at the femtometer scale. Conversely, if the hadron emission source can be estimated through other means, momentum correlation measurements allow us to probe $|\Psi^{(-)}_{\bm{q}}(\bm{r})|^{2}$, which contains information about the two-body interaction.

In correlation function measurements, since the hadrons produced in the collision are utilized, there is no need to prepare a target as in traditional scattering experiments. Moreover, even if the momenta of the individual particles $\bm{p}_1$ and $\bm{p}_2$ are large, hadron pairs with small relative momentum $\bm{q}$ correspond to low-energy scattering, making them suitable for extracting the hadron-hadron interactions. A significant advantage of high-energy proton-proton and heavy-ion collision experiments is the large number of hadrons produced per event, as mentioned above.  

In typical analyses, a spherically symmetric source function $S(r)$ with $r=|\bm{r}|$ is often assumed. In this case, the Koonin-Pratt formula gives the one-dimensional correlation function $C(q)$ with $q=|\bm{q}|$. However, in reactions where the anisotropy of the source shape becomes important, such as in heavy-ion collisions, this approach requires careful consideration~\cite{Morita:2019rph}. Furthermore, while Eq.~(\ref{eq:KPformula}) is obtained by integrating the time dependence of hadron pair production in the rest frame of the hadron pair~\cite{ExHIC:2017smd}, it has been pointed out that when hadron resonances with finite lifetimes are produced during the reaction, their contributions would effectively enlarge the source size. In fact, for the two-baryon system in $pp$ collisions, the contribution of resonance decays has been investigated, and it has been shown that the data can be described with a common spherical source function that adds a tail representing the resonance decay to a Gaussian core~\cite{ALICE:2020ibs}.

\begin{figure}[tb]
  \centering
  \includegraphics[width=7.5cm]{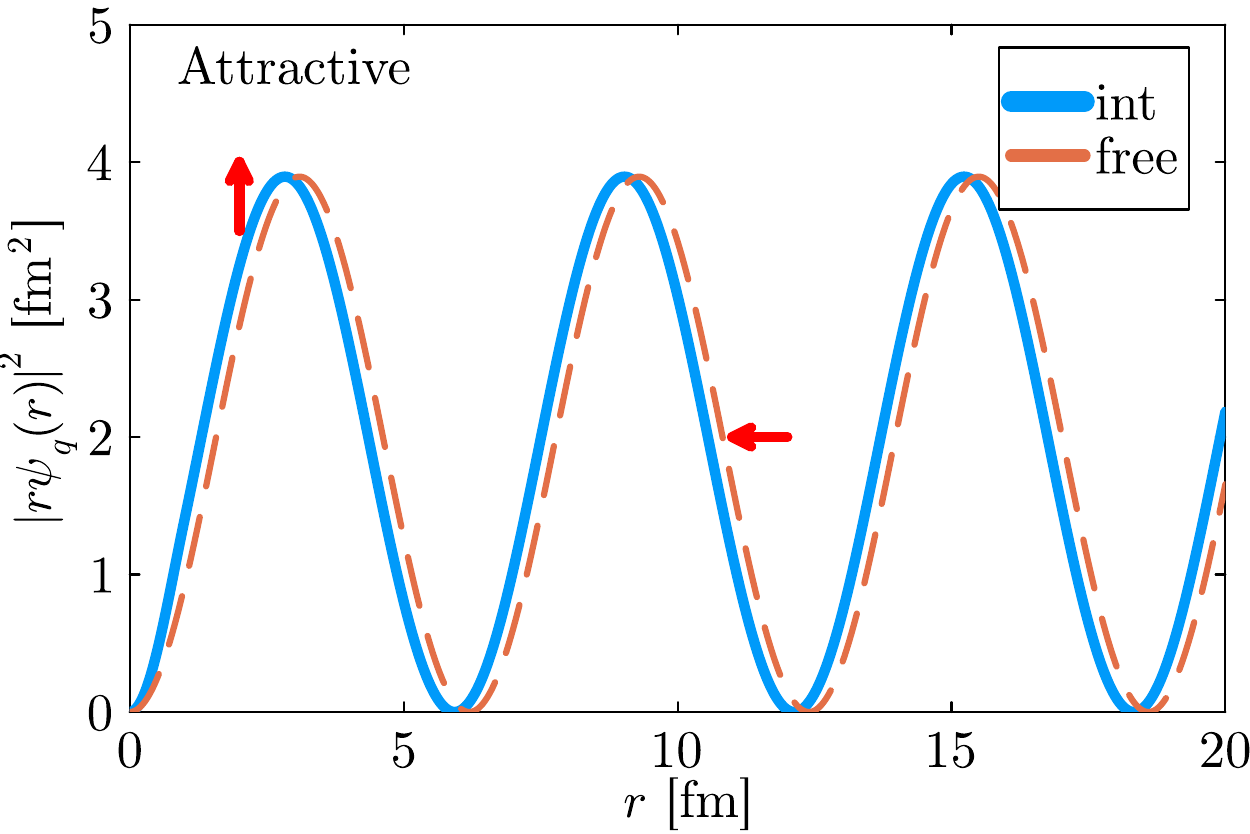}
  \includegraphics[width=7.5cm]{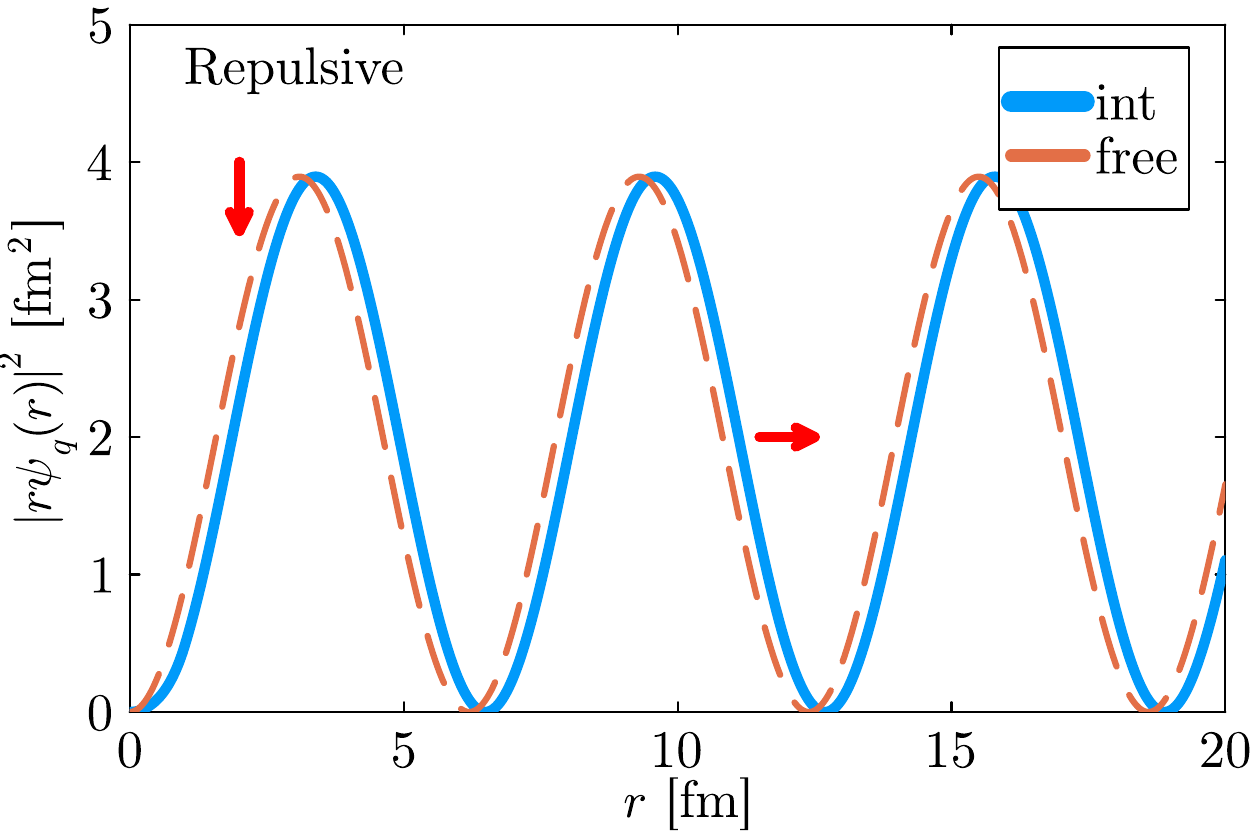}
  \includegraphics[width=7.5cm]{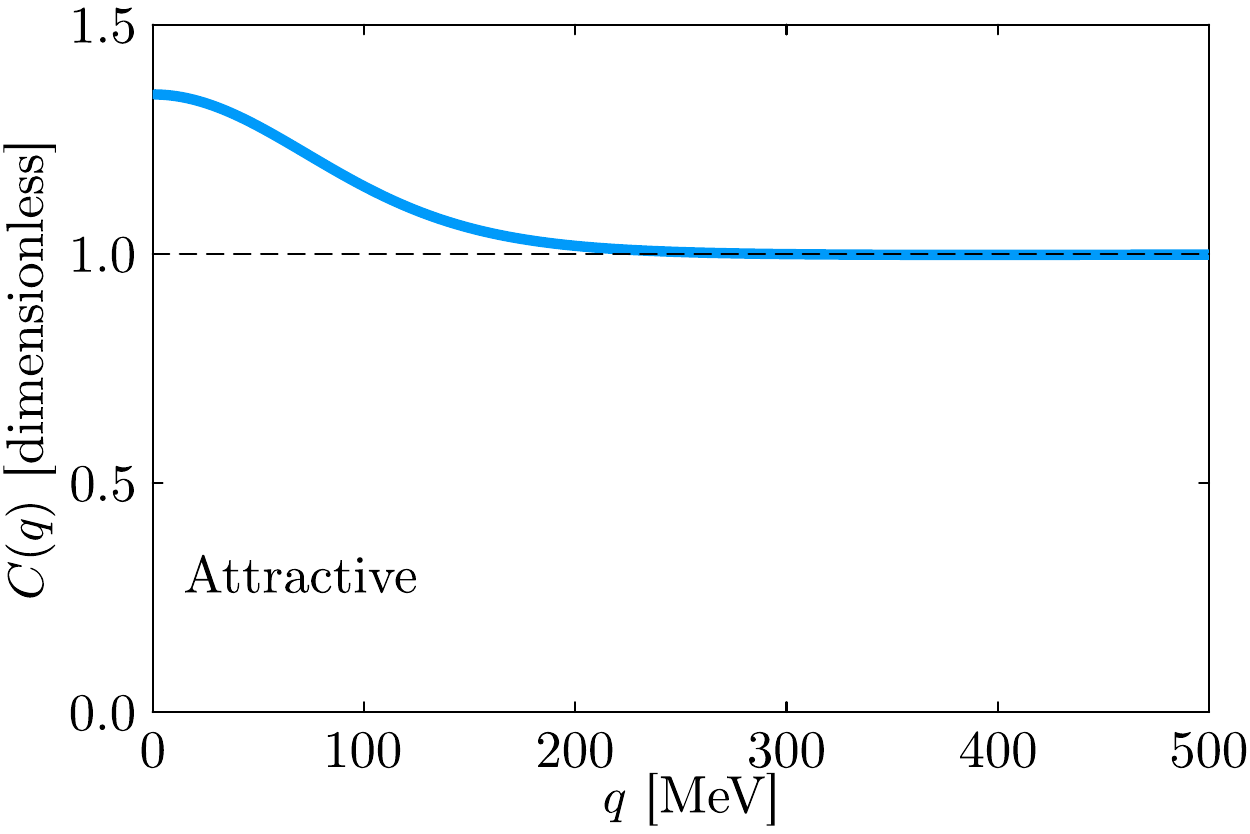}
  \includegraphics[width=7.5cm]{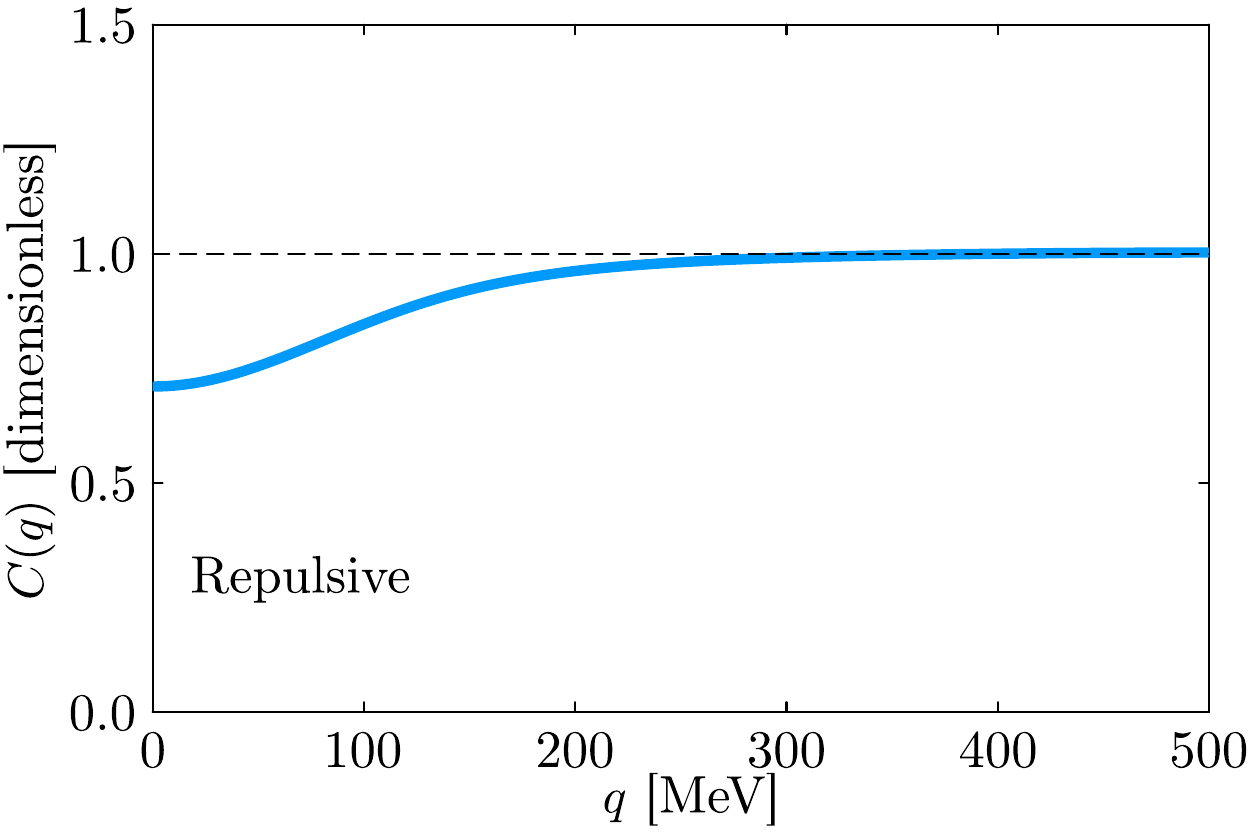}
  \caption{Wave function squared $|r\psi_{q}(r)|^{2}$ with $q=100$ MeV (upper panels) and the correlation functions (lower panels) for the square well potential. Left (right) panels show the results with weakly attractive (repulsive) interaction. }
  \label{fig:Correlation}
\end{figure}

Let us examine the behavior of the correlation function using simple interactions. Assuming a spherically symmetric source and that the interaction acts only in the $s$-wave, the correlation function is expressed as~\cite{Morita:2014kza,ExHIC:2017smd,Murase:2024ssm}  
\begin{align}
   C(q) &\simeq  
   1+4\pi\int_{0}^{\infty} dr\
   S(r)\{|r\psi_{q}(r)|^{2}-[rj_{0}(qr)]^{2}\} ,
   \label{eq:KPswave}
\end{align}
where $\psi_{q}(r)$ is the $s$-wave component of the scattering wave function with momentum $q$ and $j_{0}$ is the spherical Bessel function. This expression indicates that the deviation of the correlation function from unity is given by the difference between the interacting wave function $\psi_{q}(r)$ and the free wave function $j_{0}(qr)$. Taking the reduced mass as $\mu=470$ MeV, corresponding to the nuclear force, we calculate the wave function at $q=100$ MeV using the square-well potential $V(r)=V_{0}\Theta(1\ {\rm fm}-r)$. The results are shown in Fig.~\ref{fig:Correlation} in comparison with the free wave functions. In the case of weak attraction ($V_0=-27$ MeV, left panel), the wave function is pulled inward, whereas in the case of repulsion ($V_0=58$ MeV, right panel), the wave function is pushed outward. By using a spherically symmetric Gaussian source function $S(r) = \exp\{-r^2/(4R^2)\}/(4\pi R^2)^{3/2}$ with $R = 1$ fm, the correlation function is calculated as shown in the lower panels of the figure. The source function is finite only in the small $r$ region, so the integral is mainly influenced by the difference of the wave functions in this region. In the case of $q=100$ MeV shown in the figure, the correlation function is enhanced (suppressed) from unity, reflecting the increase (decrease) of the wave function. When $q$ becomes sufficiently large, the wavelength shortens, causing the integrand to oscillate rapidly between positive and negative values. As a result, the deviation of the correlation function from unity disappears. In this way, the correlation function qualitatively reflects the nature of the interaction. It is also known that the correlation function exhibits a strong dependence on the source size, when the attraction is strong enough to support a bound state, due to the node of the wave function~\cite{ExHIC:2017smd,Kamiya:2021hdb}.

\section{Femtoscopy for exotic hadrons}

One of the key applications of femtoscopy is in the study of the $\bar{K}N$ interactions, which play a crucial role in understanding the nature of the $\Lambda(1405)$ resonance. The $\Lambda(1405)$ resonance is difficult to describe within the conventional three-quark picture of the quark model~\cite{Isgur:1978xj}, and it is expected to have an exotic internal structure. In particular, given that its mass is close to the $\bar{K}N$ threshold and that it decays into $\pi\Sigma$ via the strong interaction, the coupling with meson-baryon components should be important. This has led to significant interest in a $\bar{K}N$ molecular structure based on the framework of coupled-channel scattering incorporating chiral SU(3) symmetry~\cite{Kaiser:1995eg,Oset:1997it,Oller:2000fj,Hyodo:2007jq,Hyodo:2011ur,Kamiya:2015aea,Kamiya:2016oao,Meissner:2020khl,Mai:2020ltx,Hyodo:2020czb,Hyodo:2022xhp,Sadasivan:2022srs}.

The application of the femtoscopy technique to the $\bar{K}N$ channel can be found in Ref.~\cite{Acharya:2019bsa}, where the $K^{-}p$ correlation function has been measured in proton-proton collisions at $\sqrt{s} = 13$~TeV by the ALICE collaboration. In the actual measurement, data from both the $K^{-}p$ pairs and the antiparticle pairs, $K^{+}\bar{p}$, are combined. We plot the data as points with error bars in Fig.~\ref{fig:Correlation}, demonstrating the high-statistical precision even at very small momentum. What is notable in the data is the non-monotonic behavior of the correlation function at a relative momentum $q\sim 58$~MeV. This corresponds to the threshold energy of $\bar{K}^0n$ ($K^0\bar{n}$) and the behavior of the correlation is interpreted as the effect of the threshold cusp. The difference in threshold energies between $K^{-}p$ and $\bar{K}^0n$ is due to the breaking of isospin symmetry, and it is only about 5 MeV in the center-of-mass energy. In this energy region, only limited and low-precision scattering data had been available, so the $\bar{K}^0n$ cusp was not visible in previous scattering data. The correlation function data have sufficient precision to confirm the $\bar{K}^0n$ threshold cusp, and since data exist in an energy region even below the $\bar{K}^0n$ threshold, it is expected to provide new constraints on the low-energy $\bar{K}N$ interaction.

For the theoretical calculation of correlation functions involving channel coupling, such as the $K^{-}p$ correlation function, the extended Koonin-Pratt formula
\begin{align}
   C(\bm{q}) &= 
   \int d^3r\  \sum_{j}\omega_{j}
   S_{j}(\bm{r})\left|\Psi_{j,\bm{q}}^{(-)}(\bm{r})\right|^{2} ,
   \label{eq:KPLLLformula}
\end{align}
should be used~\cite{Lednicky:1997qr,Haidenbauer:2018jvl,Kamiya:2019uiw}. Here, $\Psi_{j,\bm{q}}^{(-)}(\bm{r})$ represents the wave function component for channel $j$, and $S_{j}(\bm{r})$ is the source function for channel $j$. The parameter $\omega_{j}$ represents the relative weight of channel $j$. In the specific case of calculating the $K^{-}p$ correlation function, $\omega_{K^{-}p} = 1$, and the relative production rate of channel $j$ for $K^{-}p$ production is represented by $\omega_{j}$. Each $j$ in Eq.~(\ref{eq:KPLLLformula}) corresponds to a channel-transition process; the channel $j$ is initially produced, turns into $K^{-}p$ through the final state interaction, and is then detected. 

In Ref.~\cite{Kamiya:2019uiw}, the $K^{-}p$ correlation function is calculated using the Kyoto $\bar{K}N$-$\pi\Sigma$-$\pi\Lambda$ potential~\cite{Miyahara:2018onh} based on chiral SU(3) dynamics~\cite{Ikeda:2011pi,Ikeda:2012au}. To perform theoretical calculations that match the precision of ALICE data, it is essential to properly incorporate the effects of channel coupling, isospin symmetry breaking, and Coulomb interactions. The source function can be determined using the correlation function of $K^{+}p$ ($K^-\bar{p}$). Since the interaction in the $K^{+}p$ system is relatively well known, the source function for this system can be estimated from the correlation function measurement. Furthermore, in the collision conditions under consideration, the production mechanisms for $K^{+}p$ and $K^{-}p$ are expected to be similar, so the source function for $K^{-}p$ can be estimated using information from the $K^{+}p$ source function. The weight parameter $\omega_{i}$ is estimated by the simple statistical model~\cite{Andronic:2005yp}. As shown in Fig.~\ref{fig:Correlation}, the result of the correlation function with the source size of $R = 0.9$ fm (solid line) shows good agreement with experimental data. A small peak around $q\sim 240$ MeV is due to the $\Lambda(1520)$ resonance, whose contribution can be parametrized by the Breit-Wigner form with the PDG values of the mass and width (dotted line). 

\begin{figure}[tb]
  \centering
  \includegraphics[width=9cm]{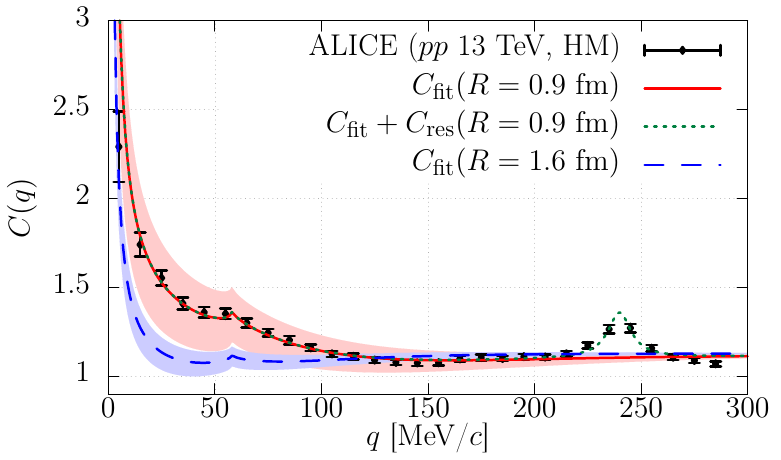}
  \caption{$K^{-} p$ correlation functions for $pp$ collisions at $\sqrt{s} =13$~TeV (solid line). Experimental data (points with errors) are taken from Ref.~\cite{Acharya:2019bsa}. The dotted line shows the result including the $\Lambda(1520)$ resonance contribution. The dashed line represents the prediction of the correlation function from larger size source. Figure adapted from Ref.~\cite{Kamiya:2019uiw}.}
  \label{fig:Correlation}
\end{figure}

Theoretically, the correlation function for cases with larger source sizes, such as those in heavy-ion collisions, can be predicted. As shown by the dashed line in Fig.~\ref{fig:Correlation}, increasing the source size to $R = 1.6$ fm suppresses the correlation function at low momenta. It is also found that the contribution from channel coupling becomes less significant~\cite{Kamiya:2019uiw}. This trend is consistent with experimental results from Pb-Pb collisions by the ALICE collaboration~\cite{ALICE:2021szj}, which produce an emission source of approximately $R = 5.2$ fm. A systematic study of the source size dependence is conducted in Ref.~\cite{ALICE:2022yyh} using data from $pp$ collisions, $p$-Pb collisions, and Pb-Pb collisions with various centralities. By utilizing coupled-channel wave functions obtained from the Kyoto potential~\cite{Kamiya:2019uiw} and weight factors $\omega_{i}$ estimated using the Thermal-FIST package~\cite{Vovchenko:2019pjl}, theoretical correlation functions are predicted. Comparison with experimental data indicates that the coupling strength to the $\bar{K}^{0}n$ channel is insufficient, suggesting the need for further refinement of the $\bar{K}N$ potential.

\section{Femtoscopy for hypernuclei}

Femtoscopic study including atomic nuclei, particularly the $\Lambda \alpha$ system, provide a unique opportunity to investigate the physics of the hypernuclei. One of the major issues in hypernuclear study is the softening of the equation of state in neutron stars due to the appearance of hyperons, known as the hyperon puzzle~\cite{Gal:2016boi,Burgio:2021vgk}. A solution to the hyperon puzzle is proposed in the framework of chiral EFT, where the $\Lambda NN$ three-body force acts repulsively at high densities, suppressing the appearance of $\Lambda$ hyperons~\cite{Gerstung:2020ktv}. As an experimental approach to test this scenario, methods such as utilizing the $\Lambda$ directed flow in heavy-ion collisions have been proposed \cite{Nara:2022kbb}. Femtoscopic studies of the correlations between $\Lambda$ and nuclei can also contribute to this issue. In general, density effects can be investigated using heavy nuclei, while producing such nuclei in high-energy collisions is challenging. Here, we focus on the $\Lambda\alpha$ correlations with two-body treatment of the $\Lambda\alpha$ system, which may be justified by the strong binding energy of the $\alpha$ particle. Also, a bunch of $\alpha$ particles may be produced in the heavy ion collisions with $\sqrt{s_{NN}}<$ 10 GeV~\cite{Andronic:2010qu}. In recent years, femtoscopy studies of interactions between light nuclei and hadrons have been conducted both experimentally~\cite{ALICE:2023bny} and theoretically~\cite{Haidenbauer:2020uew,Ogata:2021mbo,Viviani:2023kxw,Kamiya:2024diw,Kohno:2024tkq}.
 
\begin{figure}[tb]
  \centering
  \includegraphics[width=7cm]{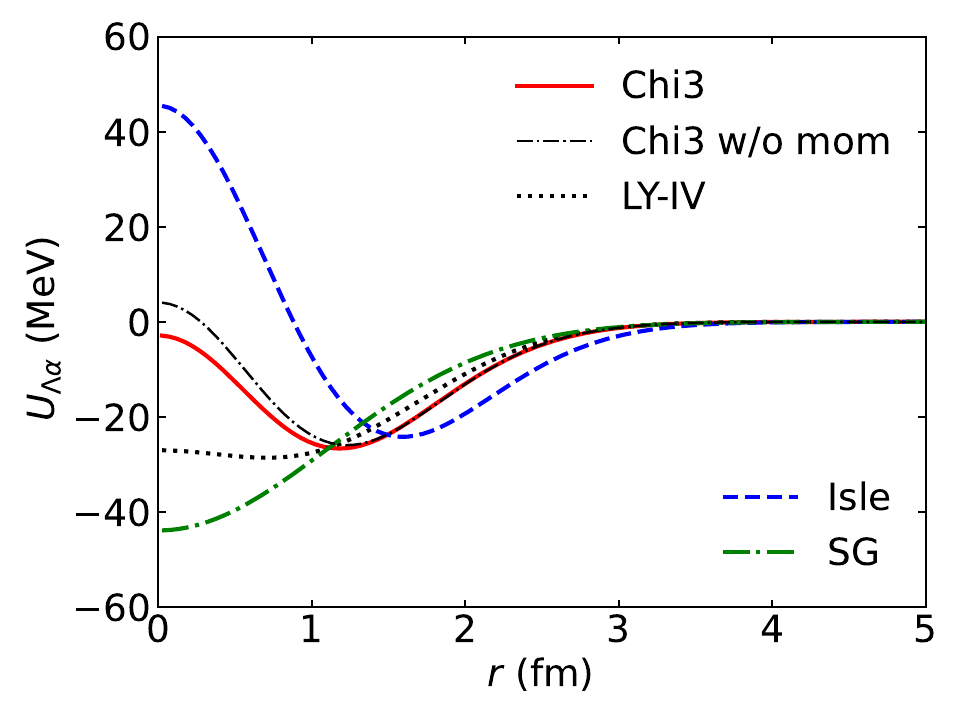}
  \includegraphics[width=7cm]{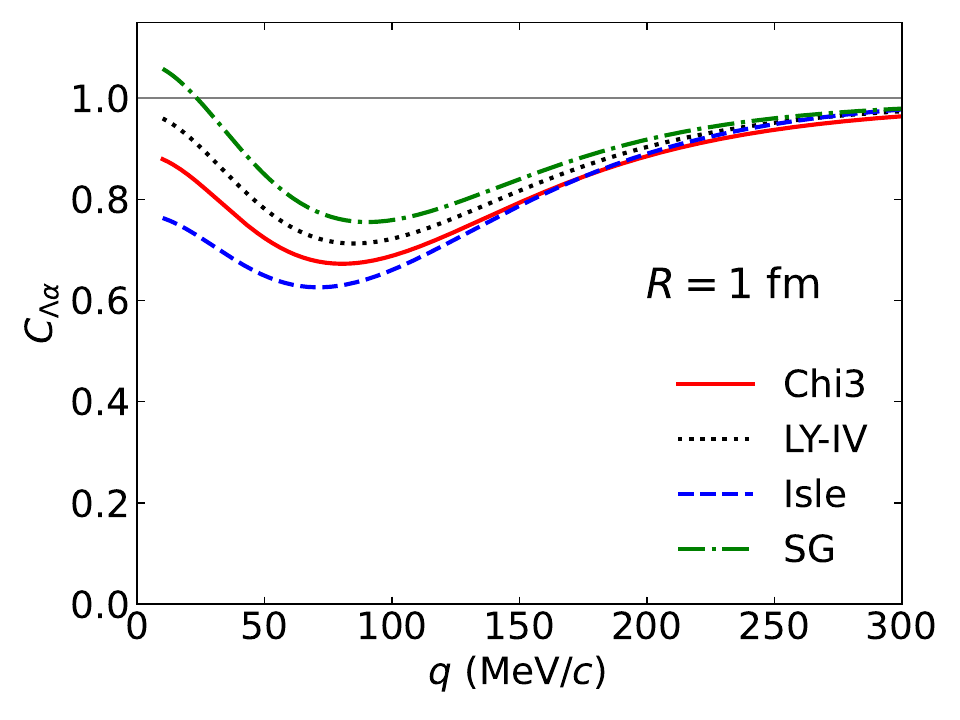}
  \caption{$\Lambda\alpha$ potentials (left) and the corresponding correlation functions (right). Figure adapted from Ref.~\cite{Jinno:2024tjh}.}
  \label{fig:Lalpha}
\end{figure}

Several models have been proposed for the $\Lambda\alpha$ potential. A phenomenological single Gaussian (SG) potential and the Isle potential having a repulsive core at short distances, have been constructed in Ref.~\cite{Kumagai-Fuse:1994ulj} with the constraint of the binding energy of ${}^{5}_{\Lambda}{\rm He}$. The $\Lambda\alpha$ potential can also be derived more microscopically by using the density functional derived from the Skyrme-Hartree-Fock method, which reproduces hypernuclear data from C to Pb, together with the density distribution of the $\alpha$ particle. In this study, we construct the potential using the phenomenological LY-IV model~\cite{Lanskoy:1997xq} and using the Chi3 model~\cite{Jinno:2023xjr} which is based on the chiral EFT results of Ref.~\cite{Gerstung:2020ktv} including the effects of the $\Lambda NN$ three-body force. The obtained potentials are shown in the left panel of Fig.~\ref{fig:Lalpha}. Although all the potentials are constructed to reproduce the binding energy of ${}^{5}_{\Lambda}{\rm He}$, they exhibit different behaviors in the short-range repulsive core. In particular, comparing LY-IV and Chi3 potentials, we find that the LY-IV potential maintains a constant attraction even in the small $r$ region, whereas the Chi3 potential develops a repulsive component at the origin. This can be interpreted as the effect of the $\Lambda NN$ three-body force, causing the $\Lambda$ to experience repulsion in the high density region near the center of the $\alpha$ particle.  

To examine the consequence of the differences in the potentials, we calculate the $\Lambda\alpha$ correlation function using these potentials~\cite{Jinno:2024tjh}. The results for the correlation functions, obtained using a Gaussian source with $R=1$ fm, are shown in the right panel of Fig.~\ref{fig:Lalpha}, indicating that different potentials yield different correlations. In particular, while the potential strengths at $r=0$ are ordered as  
\begin{align}
   U_{\rm Isle}(r=0) > U_{\rm Chi3}(r=0) > U_{\rm LY-IV}(r=0) > U_{\rm SG}(r=0), 
\end{align}
the corresponding correlation functions follow the order  
\begin{align}
   C_{\rm Isle} < C_{\rm Chi3} < C_{\rm LY-IV} < C_{\rm SG}.
\end{align}
This result demonstrates that potentials with a stronger repulsive core near the center have a greater suppression of the correlation function from unity. In this way, we find that the femtoscopy technique enables the experimental investigation of the repulsive core of the $\Lambda\alpha$ potential through precise correlation function measurements.

\section{Summary}

Femtoscopy has emerged as a crucial tool in studying hadronic interactions. Since high-precision data can be obtained for hadronic systems that are difficult to study in conventional scattering experiments, novel studies on hadron-hadron interactions become possible. The precise measurements of $K^-p$ correlations provide stringent tests for the $\bar{K}N$ potential based on chiral SU(3) dynamics and the nature of the $\Lambda(1405)$ resonance. The correlation function for $\Lambda \alpha$ pairs is sensitive to the nature of the $\Lambda \alpha$ potential, in particular to the presence of a repulsive core. Future femtoscopic experiments at high-energy colliders will further refine our understanding of exotic hadrons and hypernuclei.

%
%

\section*{Acknowledgments}

The author thanks Ayse Kizilersu and Nora Brambilla for the kind invitation to the wonderful conference. The author expresses gratitude to all collaborators involved in the research presented in this contribution. In particular, the author would like to extend special thanks to the late Akira Ohnishi, who played a central role in initiating the theoretical femtoscopy approach for hadron-hadron interactions, but sadly passed away in May 2023. Around 2015, Akira invited the author to explore femtoscopy research on the $K^-p$ correlations together. The opportunity to conduct studies on the $K^-p$ and $\Lambda\alpha$ correlations presented here, in collaboration with Akira, has been an invaluable experience for the author. The author wishes to express his deepest condolences on the passing of Akira Ohnishi.

%

%

\end{document}